\documentclass[twocolumn,prl,superscriptaddress,amsmath,amssymb]{revtex4-1}
\usepackage{graphicx,bm}
\usepackage{color}
\usepackage[colorlinks=true,citecolor=blue,urlcolor=blue]{hyperref}
\usepackage{soul}
\usepackage{siunitx}

\begin{document}
	
\title{Light Manipulation via Tunable Collective Quantum States in Waveguide-Coupled Bragg and Anti-Bragg Superatoms}

\author{Zhengqi Niu}
\affiliation{State Key Laboratory of Materials for Integrated Circuits, Shanghai Institute of Microsystem and Information Technology, Chinese Academy of Sciences, Shanghai 200050, China}
\affiliation{ShanghaiTech University, Shanghai 201210, China}
	
\author{Wei Nie}\email{weinie@tju.edu.cn}
\affiliation{Center for Joint Quantum Studies and Department of Physics, School of Science, Tianjin University, Tianjin 300350, China}
\affiliation{Tianjin Key Laboratory of Low Dimensional Materials Physics and Preparing Technology, Tianjin University, Tianjin 300350, China}
	
\author{Daqiang Bao}
\affiliation{State Key Laboratory of Materials for Integrated Circuits, Shanghai Institute of Microsystem and Information Technology, Chinese Academy of Sciences, Shanghai 200050, China}

\author{Xiaoliang He}
\affiliation{State Key Laboratory of Materials for Integrated Circuits, Shanghai Institute of Microsystem and Information Technology, Chinese Academy of Sciences, Shanghai 200050, China}
\affiliation{University of Chinese Academy of Sciences, Beijing 100049, China}
	
\author{Wanpeng Gao}
\affiliation{State Key Laboratory of Materials for Integrated Circuits, Shanghai Institute of Microsystem and Information Technology, Chinese Academy of Sciences, Shanghai 200050, China}
\affiliation{University of Chinese Academy of Sciences, Beijing 100049, China}
	
\author{Kuang Liu}
\affiliation{State Key Laboratory of Materials for Integrated Circuits, Shanghai Institute of Microsystem and Information Technology, Chinese Academy of Sciences, Shanghai 200050, China}
	
\author{I.-C. Hoi}
\affiliation{Department of Physics, City University of Hong Kong, Kowloon, Hong Kong SAR 999077, China}
	
\author{Yu-xi Liu}
\email[]{yuxiliu@mail.tsinghua.edu.cn}
\affiliation{School of Integrated Circuits,Tsinghua University, Beijing 100084, China}
\affiliation{Frontier Science Center for Quantum Information, Beijing, China}
	
\author{Xiaoming Xie}
\affiliation{State Key Laboratory of Materials for Integrated Circuits, Shanghai Institute of Microsystem and Information Technology, Chinese Academy of Sciences, Shanghai 200050, China}
\affiliation{University of Chinese Academy of Sciences, Beijing 100049, China}
	
\author{Zhen Wang}
\affiliation{State Key Laboratory of Materials for Integrated Circuits, Shanghai Institute of Microsystem and Information Technology, Chinese Academy of Sciences, Shanghai 200050, China}
\affiliation{ShanghaiTech University, Shanghai 201210, China}
\affiliation{University of Chinese Academy of Sciences, Beijing 100049, China}
	
\author{Zhi-Rong Lin}
\email[]{zrlin@mail.sim.ac.cn}
\affiliation{State Key Laboratory of Materials for Integrated Circuits, Shanghai Institute of Microsystem and Information Technology, Chinese Academy of Sciences, Shanghai 200050, China}
\affiliation{University of Chinese Academy of Sciences, Beijing 100049, China}
	
\begin{abstract}
A many-body quantum system which consists of collective quantum states, such as superradiant and subradiant states, behaves as a multi-level superatom in light-matter interaction. In this work, we experimentally study one-dimensional superatoms in waveguide quantum electrodynamics with a periodic array of superconducting artificial atoms. We engineer the periodic atomic array with two distinct nearest-neighbor spacings, i.e., $d$=$\lambda_0/2$ and $d$=$\lambda_0/4$, which correspond to Bragg and anti-Bragg scattering conditions, respectively. The system consists of eight atoms arranged to maintain these specific interatomic distances. By controlling atomic frequencies, we modify Bragg and anti-Bragg superatoms, resulting in distinctly different quantum optical phenomena, such as collectively induced transparency and a broad photonic bandgap. Moreover, due to strong waveguide-atom couplings in superconducting quantum circuits, efficient light manipulations are realized in small-size systems. Our work demonstrates tunable optical properties of Bragg and anti-Bragg superatoms, as well as their potential applications in quantum devices.
\end{abstract}
	
\maketitle
	
\textit{Introduction}.---Electromagnetic fields can induce collective quantum behaviours between atoms, such as enhanced (suppressed) spontaneous emission, i.e., superradiance (subradiance)~\cite{PhysRev.93.99,gross1982,PhysRevLett.101.153601,scully2009,lodahl2017chiral,RevModPhys.90.031002,PRXQuantum.3.010201}, and cooperative Lamb shifts~\cite{PhysRevLett.102.143601,rohlsberger2010,PhysRevLett.113.193002,PhysRevLett.123.233602,hutson2024}. The cooperativity makes multiple atoms an effective superatom~\cite{vuletic2006,PhysRevX.7.041010,PhysRevLett.124.023603,PhysRevLett.125.073602}, consisting of subradiant and superradiant states. Quantum interference between these collective states is important for spontaneous radiation process, and can produce novel quantum effects, e.g., collectively induced transparency (CIT)~\cite{lei2023}, directional photon absorption and emission~\cite{PhysRevApplied.15.044041,PhysRevLett.128.113601}. Optical properties of superatoms depend on microscopic interactions between atoms and electromagnetic fields. For example, in waveguide quantum electrodynamics (QED), one-dimensional (1D) continuum induces long-range dipole-dipole interactions~\cite{PhysRevLett.106.020501,van2013photon,PhysRevLett.115.063601,Solano2017,Brehm2021,tiranov2023}. Spatial degrees of freedom in atom arrays are useful for light manipulation~\cite{corzo2019waveguide,kannan2020generating,PhysRevLett.128.073601,RevModPhys.89.021001,RevModPhys.95.015002,gonzalez2024light}, and enable broad applications in waveguide-based quantum devices, such as quantum memories~\cite{PhysRevX.7.031024,Mirhosseini2019,PhysRevLett.122.203605,Zanner2022}.
	
Photon transport in periodic structures is of great interest in artificial quantum materials, e.g., photonic crystals~\cite{PhysRevLett.58.2059,PhysRevLett.65.3152,joannopoulos1997} and optical lattices~\cite{PhysRevA.52.1394,PhysRevLett.75.2823,PhysRevLett.106.223903}. In order to enhance light reflection in these systems, Bragg condition is required. In a waveguide coupled with an atom array, the Bragg condition means that the spacing $d$ between two nearest-neighboring atoms should be $n\lambda_0/4$ where $n$ is an even number~\cite{Chang2012cavity}. Bragg scattering has been experimentally observed in 1D atom arrays trapped by optical nanofibers~\cite{PhysRevLett.117.133603,PhysRevLett.117.133604}. However, due to low coupling factor $\beta$, which is defined as the ratio of decay rate into the guided mode to total decay rate of the individual atom~\cite{PhysRevB.75.205437,PhysRevLett.99.023902}, hundreds of atoms are needed to realize strong reflection~\cite{PhysRevLett.109.033603}. Different from Bragg atom arrays, anti-Bragg scattering occurs at the atomic spacing $d$=$n\lambda_0/4$ with an odd number $n$~\cite{PhysRevLett.76.4199}, and receives growing attention in waveguide QED~\cite{PhysRevApplied.15.044041,PhysRevLett.131.103602,PhysRevA.106.L031702,PhysRevA.110.053716,PhysRevA.111.023707}. However, due to the challenge to precisely control atomic positions, optical properties of anti-Bragg atom arrays have not been experimentally demonstrated yet.

\begin{figure*}[t]
	\centering
	\includegraphics[width=17.5cm]{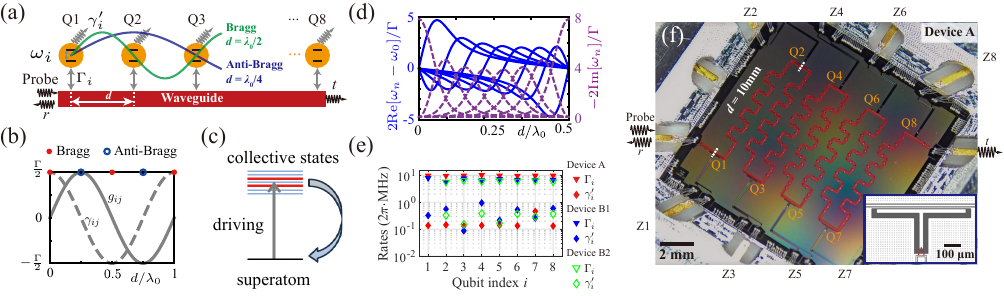}
	\caption{(a) The schematics of a waveguide QED system with $N=8$ qubits. A periodic qubit array (labelled as Q\textit{i}) is coupled to an open one-dimensional waveguide with spacing $d$. For the \textit{i}th qubit, $\omega_i$ is the transition frequency, ${\rm\Gamma}_i$ is the waveguide-induced decay rate, $\gamma_i'$ is the dissipation to free space. A probe field is sent to the waveguide from left. Transmitted and reflected signals are denoted by \textit{t} and \textit{r}, respectively. (b) Waveguide-mediated coherent ($g_{ij}$) and dissipative ($\gamma_{ij}$) couplings between Q\textit{i} and Q\textit{j} are changed as a function of $d/\lambda_0$. The red and blue dots indicate the Bragg and anti-Bragg conditions with vanishing waveguide-mediated coherent and dissipative coupling, respectively. (c) The schematic diagram of photon emission process in a driven superatom with collective quantum states. (d) Real and imaginary parts of spectrum $\omega_n$ as a function of $d/\lambda_0$. (e) Measured waveguide-induced decay rates $\Gamma_i$ and dissipation to free space $\gamma_i'$ in Device A, B1 and B2. (f) The optical image of Device A. Qubits are labelled by Q\textit{i}. The waveguide is indicated by red winding line. Each qubit is equipped with an individual flux line, labelled as Z\textit{i}. Inset panel shows the layout of the transmon qubit near the waveguide with flux line.}\label{fig.1}
\end{figure*}
	
In this Letter, we experimentally study waveguide QED with 1D superatoms in superconducting circuits. With well-controlled distances between superconducting artificial atoms or qubits in waveguides, fundamental optical properties of Bragg and anti-Bragg superatoms are demonstrated. Different from single atoms~\cite{Astafiev2010,PhysRevLett.107.073601,kannan2020waveguide}, spatial degrees of freedom provide versatile tunability in quantum coherence properties of superatoms. We show that frequencies of individual atoms nontrivially change quantum interference between collective quantum states, producing CIT effects in Bragg-type superatoms, and a broad photonic bandgap in anti-Bragg configurations. Remarkably, owing to high $\beta$ factor, these novel collective quantum phenomena emerge in minimal systems comprising just a few superconducting artificial atoms.

\textit{1D superatoms in waveguide quantum systems}.---Figure \ref{fig.1}(a) shows the schematics of a waveguide QED system with $N=8$ qubits. The system can be described by an effective non-Hermitian Hamiltonian ($\hbar=1$)
\begin{equation}
H_{\mathrm{eff}} =\sum_{i=1}^N \left(\omega_i - i\frac{\gamma'_i}{2}\right) \sigma_i^+\sigma_i^- +  \sum_{i,j=1}^N \left(g_{ij} -i \gamma_{ij} \right) \sigma_i^+ \sigma_j^-, \label{Eq1}
\end{equation}
where $\omega_i$ is the frequency of the $i$th qubit, $\gamma'_i$ is its dissipation to free space. Homogeneous frequencies $\omega_i=\omega_0$ can be realized in superconducting qubits. For superconducting waveguide QED systems, waveguide-mediated coherent and dissipative couplings are $g_{ij}=\sqrt{\Gamma_i \Gamma_j}\sin(2\pi d_{ij}/\lambda_0)/2$ and $\gamma_{ij}=\sqrt{\Gamma_i \Gamma_j}\cos(2\pi d_{ij}/\lambda_0)/2$~\cite{PhysRevA.88.043806}, respectively. Here, $\Gamma_{i}$ is waveguide-induced decay rate of the $i$th qubit, $d_{ij}=|x_i - x_j|$ is the distance between $i$th and $j$th qubits, $\lambda_0$ is the wavelength of a photon with frequency $\omega_0$. Positions of qubits in a waveguide can be well-controlled in superconducting quantum circuits. This allows us to study optical properties of periodic qubit arrays. As shown in Fig.~\ref{fig.1}(b), Bragg (anti-Bragg) qubit arrays have vanishing waveguide-mediated nearest-neighboring coherent (dissipative) couplings. These two types of periodic qubit arrays produce distinctly different quantum interference phenomena in photon transport, i.e., Bragg and anti-Bragg photon scattering phenomena.
	
The effective non-Hermitian Hamiltonian in Eq.~(\ref{Eq1}) can be diagonalized as $H_{\mathrm{eff}}=\sum_n \omega_n |\Psi_n^R\rangle \langle \Psi_n^L|$, with the complex spectrum $\omega_n$ and biorthogonal basis $\langle\Psi_m^L|\Psi_n^R\rangle=\delta_{mn}$\cite{Brody_2014}. The index $n$ labels collective quantum states in the system. A waveguide-coupled qubit array behaves as a 1D superatom~\cite{PhysRevLett.124.023603}, whose excited states are collective quantum states. A driven-dissipative superatom gives rise to novel quantum optical phenomena due to quantum interference between collective quantum states in photon emission process, as shown in Fig.~\ref{fig.1}(c). Quantum coherence distribution in collective quantum states determines optical properties of superatoms~\cite{PhysRevLett.124.023603}. In Fig.~\ref{fig.1}(d), we show real and imaginary parts of spectrum $\omega_n$~\cite{PhysRevA.95.033818}. The spacing $d$ nontrivially changes energy levels and quantum coherence (decay rates) of the superatom. At Bragg and anti-Bragg scattering conditions, superatoms have symmetric energy levels. In particular, the Bragg-type superatom has degenerate collective states, including a single superradiant states with \textit{N}-fold enhanced decay rate $N\Gamma$ and $N-1$ dark states. Conversely, the anti-Bragg superatom manifests nondegenerate energy levels with two superradiant states and $N-2$ subradiant or dark states~\cite{PhysRevApplied.15.044041}. Spectral differences in Bragg and anti-Bragg superatoms dramatically alter photon transport in the waveguide. Here, the $\beta$ factor defined as $\Gamma_i/(\Gamma_i+\gamma'_i)$ is important for the observation of collective quantum phenomena in waveguide QED. In Fig.~\ref{fig.1}(e), we show radiative and nonradiative decay rates of qubits in our Devices A, B1 and B2~\cite{SupplementalMaterial}. With well-controlled qubit-waveguide couplings, qubits have homogeneous radiative decay rates $\Gamma$. We have average $\beta \approx 0.921$ for strong couplings in these three devices. Therefore, distinct collective quantum effects in Bragg and anti-Bragg superatoms can be observed in small-size superconducting waveguide quantum systems.

In our experiments, Bragg and anti-Bragg superatoms are implemented in superconducting quantum chips with $d=10$~mm and $d=5$~mm, which are named as Device A and B1/B2, respectively. The optical image of Device A is shown in Fig.~\ref{fig.1}(f). The superconducting coplanar waveguide (CPW) winds through the major area of the chip. Eight flux-tunable transmon qubits~\cite{PhysRevA.76.042319} are capacitively coupled to the CPW. All qubits' frequencies are tunable via individual flux lines. Bragg (anti-Bragg) condition is met approximately for qubits' frequencies $\omega_0/2\pi=6$~GHz in Device A (Devices B1 and B2). The size of capacitors is $500~{\rm\mu m}$, which is small compared with the single-photon wavelength $\lambda_0$ (e.g., $\lambda_0=2\pi v/\omega_0\approx20~\mathrm{mm}$ at $\omega_0=6$~GHz with light velocity in CPW $v\approx 1.2\times10^8$ m/s). Therefore, the qubits are point-like quantum emitters near the waveguide, such that Bragg and anti-Bragg conditions can be satisfied.

\begin{figure}[t]
\centering
\includegraphics{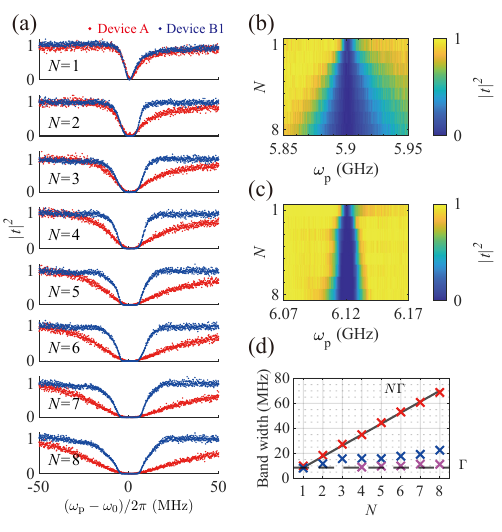}
\caption{Transmission spectra of Bragg and anti-Bragg superatoms in Devices A and B1. (a) Measured $|t|^{2}$ spectrum when the qubits are successively tuned into resonance from Q1 to Q8 in Device A ($\omega_0/2\pi=5.900$ GHz, red) and Device B1 ($\omega_0/2\pi=6.120$ GHz, blue). (b), (c) The scaled color images of $|t|^2$ spectrum depending on size $N$ of Bragg superatom in Device A and anti-Bragg superatom in Device B1. (d) Extracted width of the bandgap depending on $N$ in Device A and B1. The red (blue) cross marks represent the FWHMs of bandgap in Device A (B1). The magenta cross marks denotes the width of flat bottom of bandgap in Device B1.}\label{fig.2}
\end{figure}

\textit{Photon scattering spectroscopy of Bragg and Anti-Bragg superatoms}.---Superatoms can be probed via photon transport in a waveguide. In fact, single-photon scattering behaviors of a superatom are related to its effective non-Hermitian Hamiltonian~\cite{PhysRevA.92.053834}. Similar to single atoms, decay rates of collective quantum states characterize light-superatom interaction in a waveguide. Light transmission and reflection amplitudes can be written as multi-channel scattering forms~\cite{PhysRevApplied.15.044041}	
\begin{eqnarray}
t(\omega_{\rm p})&=&1-\frac{i\Gamma}{2}\sum_n \frac{\bm{V}^{\dagger} |\Psi_n^R\rangle  \langle \Psi_n^L| \bm{V}}{\omega_{\rm p}-\omega_n}, \label{Eqtarray} \\
r(\omega_{\rm p})&=&-\frac{i\Gamma}{2}\sum_n \frac{\bm{V}^{\mathrm{T}} |\Psi_n^R\rangle  \langle \Psi_n^L| \bm{V}}{\omega_{\rm p}-\omega_n}, \label{Eqrarray}
\end{eqnarray}
respectively, with $\bm{V}=(e^{ i k_0 x_1}, e^{i k_0 x_2}, \cdots)^{\mathrm{T}}$ and the frequency $\omega_{\rm p}$ of probe field. Collective quantum states provide different scattering channels for incident photons. Therefore, frequencies and decay rates of collective quantum states in superatoms are responsible for photon transport. To reveal differences of collective quantum states in photon scattering of Bragg and anti-Bragg superatoms, we successively tune frequencies of qubits into resonance, respectively~\cite{SupplementalMaterial}. In Fig.~\ref{fig.2}(a), we show transmittance $|t|^2$ with increased sizes \textit{N} of superatoms in Devices A and B1 . Bragg superatoms give rise to Lorentzian transmission spectra. Different from scattering phenomena demonstrated in cold atoms~\cite{PhysRevLett.117.133603,PhysRevLett.117.133604}, here we show Bragg photon scattering in consistent with theoretical studies~\cite{Chang2012cavity}, because of high $\beta$ factor and well-controlled positions of qubits. And the linewidth increases linearly with system's size. However, the anti-Bragg superatom shows different photon transport properties. Photon reflection is enhanced near the resonance frequency $\omega_0$, producing a flatband structure with vanishing transmission~\cite{PhysRevA.83.013825}. Symmetric line shapes shown in Fig.~\ref{fig.2}(a) are the evidence of Bragg and anti-Bragg scattering conditions.
	
The scaled color images of $|t|^2$ depending on the size $N$ of Bragg and anti-Bragg superatoms are shown in Figs.~\ref{fig.2}(b) and \ref{fig.2}(c), respectively. We extract the full width at half maximum (FWHM) of the spectra and plot them versus $N$ in Fig.~\ref{fig.2}(d). The red (blue) cross marks corresponds to the Bragg (anti-Bragg) superatom in Device A (B1). For Bragg superatom, the FWHM shows a clear linear dependence on \textit{N}, as evidenced by the black solid line fit in Fig.~\ref{fig.2}(d). Here, the fitted $\Gamma/2\pi$ is $8.76$~MHz. The linear scaling behavior indicates the collective superradiant state with radiative decay rate $N\Gamma$ under the Bragg condition. For anti-Bragg superatom, the optical responses show flat bandgaps as $N$ increases. 
Consequently, the anti-Bragg photonic bandgap width exhibits only weak dependence on size \textit{N}, as shown by the magenta cross markers in Fig.~\ref{fig.2}(d). The waveguide photonic bandgap comes from energy gap between two superradiant states in anti-Bragg superatom, whose width stays $\Gamma$ as \textit{N} increases~\cite{SupplementalMaterial}. These two superradiant states show constructive quantum interference in photon reflection. Therefore, input optical fields with frequencies between these two superradiant states are completely reflected.
	
\begin{figure}[t]
\centering
\includegraphics{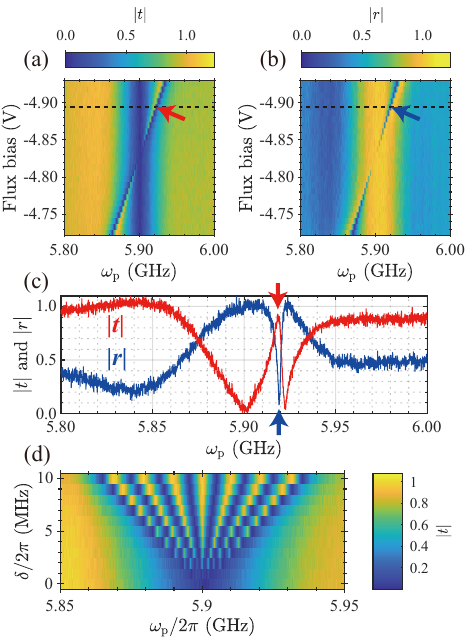}
\caption{Collectively induced transparency in the Bragg superatom. (a), (b) Transmission and reflection amplitudes $|t|$, $|r|$ versus the flux bias voltage of Q8. (c) Cross-sectional transmission (red) and reflection (blue) spectra at Q8's detuning, corresponding to arrow positions in panels (a) and (b). Arrows in (c) indicate the single collectively induced transparency window in transmission and reflection spectra. (d) The transmission amplitudes $|t|$ versus nearest-neighboring detuning $\delta$. Qubits' frequencies range from $\omega_0-3.5\delta$ to $\omega_0+3.5\delta$ with equal frequency difference $\delta$ (Q1 to Q8, $\omega_0/2\pi$ = 5.900 GHz).}\label{fig.3}
\end{figure}

\textit{Collectively induced transparency}.---Bragg and anti-Bragg superatoms exhibit different optical responses when the frequency of a single qubit is changed. Without loss of generality, we tune the last qubit (Q8) in the array. This nontrivially modifies quantum coherence of superatoms. For the Bragg superatom, Figs.~\ref{fig.3}(a) and \ref{fig.3}(b) show transmission and reflection amplitudes $|t|$, $|r|$ versus the flux bias voltage of Q8, corresponding to the Q8's frequency ranging from 5.87 {GHz} to 5.93 {GHz}. Cross-sectional transmission and reflection spectra are shown in Fig.~\ref{fig.3}(c). A transparency window is found when Q8 is slightly detuned from the array's resonance. Both the peak position (frequency) and its linewidth show systematic variations with Q8 detuning parameters. This is the collectively induced transparency (CIT) in waveguide QED~\cite{cheng2024collectively}, which is produced by quantum interference between superradiant and subradiant states~\cite{SupplementalMaterial}. The CIT effect has been observed in many-body cavity QED systems via strong driving of subradiant states~\cite{lei2023}. Here, by tuning a single qubit, we can control the frequency and decay rate of a subradiant state, producing CIT with weak driving. Frequency detuning induces the formation of a subradiant state, which subsequently mediates Fano interference in transmission spectrum. By considering homogenous frequency detunings between nearest-neighboring qubits, as shown in Fig.~\ref{fig.3}(d), we observe multi-frequency CIT~\cite{cheng2024collectively}, which indicates the generation of more subradiant states.
	
\begin{figure}[t]
\centering
\includegraphics{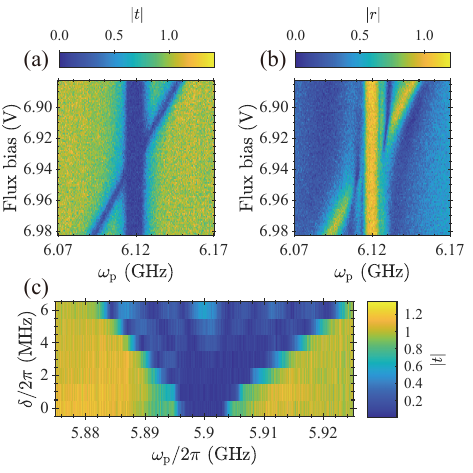}
\caption{Photonic bandgap in the anti-Bragg superatom. (a, b) The transmission and reflection amplitudes $|t|$, $|r|$ versus the flux bias voltage of Q8 in Device B1. (c) The transmission amplitudes $|t|$ versus nearest-neighboring detuning $\delta$. The qubits' frequencies range from $\omega_0-2.5\delta$ to $\omega_0+3.5\delta$ (Q2 to Q8, $\omega_0/2\pi$ = 5.900 GHz). Data in (c) were acquired using Device B2 (same design as B1 but with nonfunctional Q1), resulting in an $N=7$ anti-Bragg system. }\label{fig.4}
\end{figure}	

\textit{Broad photonic bandgap}.---Different from the Bragg counterpart, anti-Bragg superatoms do not produce CIT effect when the frequency of a single qubit is changed~\cite{SupplementalMaterial}. In Fig.~\ref{fig.4}(a) and Fig.~\ref{fig.4}(b), we show transmission and reflection by changing Q8's frequency, respectively. In anti-Bragg superatoms, there are two superradiant states. The flatband reflection of the anti-Bragg superatom shown in Fig.~\ref{fig.2}(a) is determined by the energy gap between two superradiant states. 
However, this flatband reflection has a narrow width. By changing the frequency difference between qubits, e.g., by means of direct couplings between qubits, one can realize broad photonic bandgap in an anti-Bragg superatom~\cite{PhysRevApplied.15.044041}. Similar to the Bragg superatom, we engineer homogeneous frequency detunings between nearest-neighbor qubits in the anti-Bragg configuration. This controlled detuning enlarges the spectral splitting between superradiant states, consequently generating an expanded photonic bandgap, as shown in Fig.~\ref{fig.4}(c). In the photonic bandgap, light reflection is enhanced by these two superradiant states. When the frequency splitting between the two superradiant states significantly exceeds their respective decay rates, the system exhibits suppressed light reflection.

\begin{figure}[]
\centering
\includegraphics{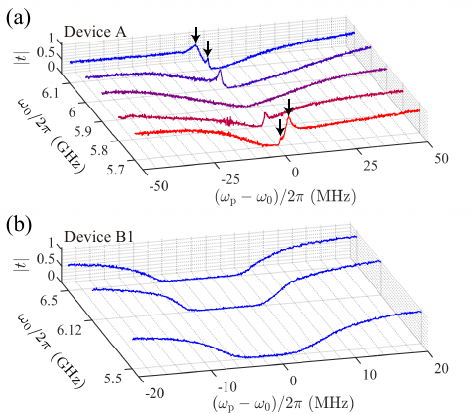}
\caption{Measured transmission amplitudes $|t|$ at several different $\omega_0$ near the Bragg (a) and anti-Bragg (b) conditions in Device A and B1, respectively. The arrows in (a) indicate multiple subradiant states observed when $\omega_0/2\pi=$ 5.700 GHz and $\omega_0/2\pi=$ 6.100 GHz.}\label{fig.5}
\end{figure}

\textit{Frequency-dependent collective quantum effects}.---Waveguide-induced couplings are controlled by spacing $d$ between nearest-neighboring qubits. This gives rise to superatoms with different energy levels and decay rates, as shown in Fig.~\ref{fig.1}(d). To investigate spacing-dependent collective quantum phenomena, we should tune the effective spacing $d/\lambda_0$. For a fixed \textit{d} in a certain device, we tune the spacing effectively by changing the resonance frequency $\omega_0$. By simultaneously modifying frequencies of qubits, we can realize qubit arrays with other effective spacings. As a result, a finite range of effective spacing $d/\lambda_0$ around $0.50$ and $0.25$ is achieved in Device A and B1, respectively. In Fig.~\ref{fig.5}(a), we show experimental measurements of transmission amplitude $|t|$ versus different frequency $\omega_0$ of qubits near the Bragg condition. The spectrum becomes asymmetric when the effective spacing $d/\lambda_0$ is changed. When the array deviates from the Bragg condition, subradiant states are produced. These subradiant states are no longer degenerate with the superradiant state (see Fig.~\ref{fig.1}(d)). Quantum interference between subradiant and superradiant states gives rise to Fano resonances. Multiple transparency windows are observed when the resonance frequency deviates far from the Bragg condition, e.g. $\omega_0/2\pi=$ 5.700 GHz and $\omega_0/2\pi=$ 6.100 GHz, as indicated by the arrows in Fig.~\ref{fig.5}(a)~\cite{SupplementalMaterial}. We also show the measured $|t|$ spectra at several different $\omega_0$ near the anti-Bragg condition in Fig.~\ref{fig.5}(b). Different from Bragg scattering, the waveguide bandgap is robust to spacings near the anti-Bragg condition.
	
\textit{Conclusions}.---In this work, we experimentally study optical properties of 1D superatoms in waveguide-integrated superconducting quantum systems. High $\beta$ factor, well-controlled positions and frequencies of superconducting qubits give rise to strong collective quantum effects in a small waveguide QED system. The spacing between nearest-neighboring qubits nontrivially modifies collective quantum states in superatoms. Distinct photon transport behaviors are observed in Bragg and anti-Bragg scattering conditions. Individual qubit frequency tuning enables the realization of two distinct quantum optical phenomena - collectively induced transparency in Bragg arrays versus broad photonic bandgap in anti-Bragg structures. Optical properties of superatoms with other spacings are also studied by simultaneously tuning frequencies of qubits. Our work shows the potential to realize light manipulation via tunable collective quantum effects in waveguide-coupled superatoms.

\begin{acknowledgments}
This project is supported by the National Key Research and Development Program of China (Grant No. 2023YFB4404904), the Strategic Priority Research Program of the Chinese Academy of Sciences (Grant No. XDB0670000) and the National Natural Science Foundation of China (Grants No. 12374483, No. 92365209, No. 62474012, No. 92476115, and No. 12105025). 
\end{acknowledgments}

\end{document}

% \preprint{123}

\title{Supplementary material for: \\"Light Manipulation via Tunable Collective Quantum States in Waveguide-Coupled Bragg and Anti-Bragg Superatoms"}

\author{Zhengqi Niu}
\affiliation{State Key Laboratory of Materials for Integrated Circuits, Shanghai Institute of Microsystem and Information Technology, Chinese Academy of Sciences, Shanghai 200050, China}
\affiliation{ShanghaiTech University, Shanghai 201210, China}

\author{Wei Nie}\email{weinie@tju.edu.cn}
\affiliation{Center for Joint Quantum Studies and Department of Physics, School of Science, Tianjin University, Tianjin 300350, China}
\affiliation{Tianjin Key Laboratory of Low Dimensional Materials Physics and Preparing Technology, Tianjin University, Tianjin 300350, China}

\author{Daqiang Bao}
\affiliation{State Key Laboratory of Materials for Integrated Circuits, Shanghai Institute of Microsystem and Information Technology, Chinese Academy of Sciences, Shanghai 200050, China}

\author{Xiaoliang He}
\affiliation{State Key Laboratory of Materials for Integrated Circuits, Shanghai Institute of Microsystem and Information Technology, Chinese Academy of Sciences, Shanghai 200050, China}
\affiliation{University of Chinese Academy of Sciences, Beijing 100049, China}

\author{Wanpeng Gao}
\affiliation{State Key Laboratory of Materials for Integrated Circuits, Shanghai Institute of Microsystem and Information Technology, Chinese Academy of Sciences, Shanghai 200050, China}
\affiliation{University of Chinese Academy of Sciences, Beijing 100049, China}

\author{Kuang Liu}
\affiliation{State Key Laboratory of Materials for Integrated Circuits, Shanghai Institute of Microsystem and Information Technology, Chinese Academy of Sciences, Shanghai 200050, China}

\author{I.-C. Hoi}
\affiliation{Department of Physics, City University of Hong Kong, Kowloon, Hong Kong SAR 999077, China}

\author{Yu-xi Liu}
\email[]{yuxiliu@mail.tsinghua.edu.cn}
\affiliation{School of Integrated Circuits,Tsinghua University, Beijing 100084, China}
\affiliation{Frontier Science Center for Quantum Information, Beijing, China}

\author{Xiaoming Xie}
\affiliation{State Key Laboratory of Materials for Integrated Circuits, Shanghai Institute of Microsystem and Information Technology, Chinese Academy of Sciences, Shanghai 200050, China}
\affiliation{University of Chinese Academy of Sciences, Beijing 100049, China}

\author{Zhen Wang}
\affiliation{State Key Laboratory of Materials for Integrated Circuits, Shanghai Institute of Microsystem and Information Technology, Chinese Academy of Sciences, Shanghai 200050, China}
\affiliation{ShanghaiTech University, Shanghai 201210, China}
\affiliation{University of Chinese Academy of Sciences, Beijing 100049, China}

\author{Zhi-Rong Lin}
\email[]{zrlin@mail.sim.ac.cn}
\affiliation{State Key Laboratory of Materials for Integrated Circuits, Shanghai Institute of Microsystem and Information Technology, Chinese Academy of Sciences, Shanghai 200050, China}
\affiliation{University of Chinese Academy of Sciences, Beijing 100049, China}

\date{\today}

% \begin{abstract}
% \end{abstract}

% %\maketitle must follow title, authors, abstract, and keywords

\setcounter{equation}{0}
\setcounter{section}{0}
\setcounter{figure}{0}
\setcounter{table}{0}
\setcounter{page}{1}

\renewcommand{\theequation}{S\arabic{equation}}
\renewcommand{\thetable}{S\arabic{table}}
\renewcommand{\thefigure}{S\arabic{figure}}

\maketitle

\clearpage

\section{measurement setup}

\begin{figure}[h]
	\centering
	\includegraphics[]{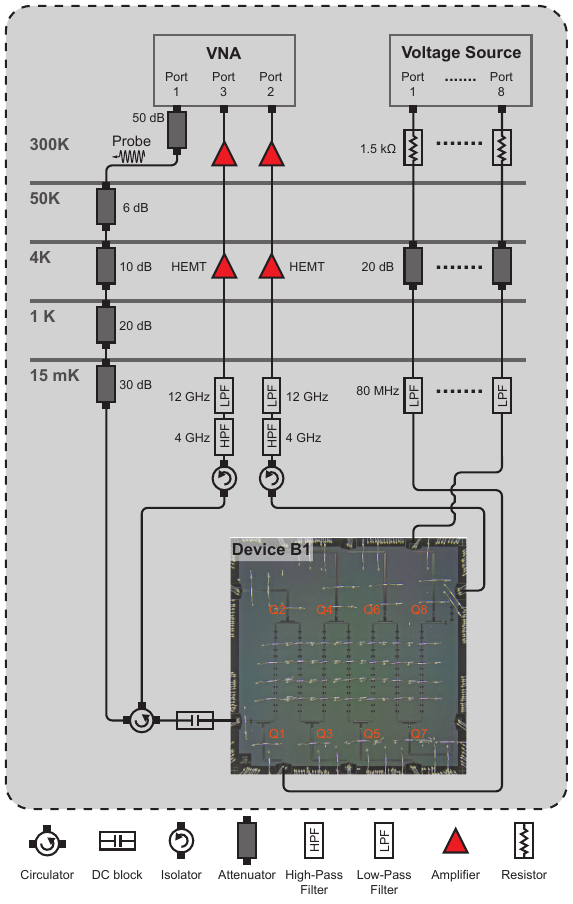}
	\caption{The schematic diagram of the experiment setup used to measure the transmission and reflection spectra while controlling the qubits' frequencies with flux lines. The optical image of Device B1 is used in this diagram for illustration. }
	\label{fig.meas}
\end{figure}

The experiments are performed in a dilution refrigerator where the samples are cooled down to the temperature of 15 mK. Fig. \ref{fig.meas} shows the schematic diagram of the measurement setup. The probe signal from the vector network analyzer (VNA) is attenuated by 50 dB at room temperature, 6 dB at the 50 K stage, 10 dB at the 4 K stage, 10 dB at the still (1 K), and 30 dB at the mixing chamber (15 mK). The sample is placed after a circulator and a DC block. The circulator is used for reflection spectrum measurement. The DC block prevents the low-frequency current noise from injecting into the waveguide. The transmitted and reflected signals go through isolators and then are filtered with 4 GHz high pass filters and 12 GHz low pass filters. The output signals are amplified with two-stage amplification, i.e., the high electron mobility transistor (HEMT) amplifiers at the 4 K stage and amplifiers at room temperature. 

The qubits' frequencies are controlled with individual flux lines. Coaxial cables are used as flux lines. Each flux line is attenuated by 20 dB at the 4 K stage and filtered by an 80 MHz low-pass filter at 15 mK stage. A set of DC voltage sources and \qty{1.5}{\kilo\ohm} resistors are used to generate the currents for frequency control. 

\clearpage

\section{single qubit characterization}

\begin{figure}[h]
\centering
\includegraphics{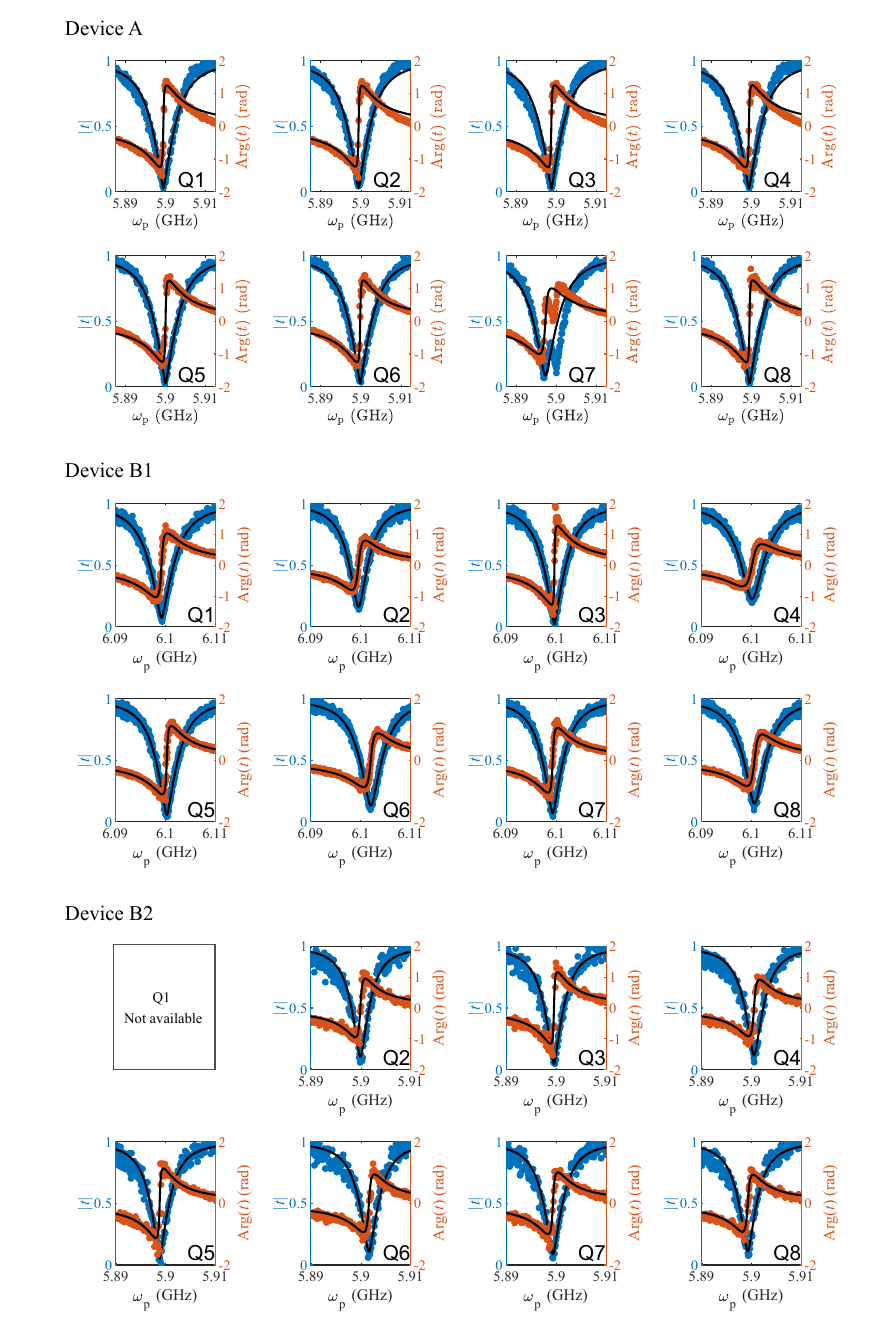}
\caption{Single qubit characterization of Devices A, B1 and B2. Each panel shows magnitude $|t|$ (left \textit{y}-axis) and phase Arg(\textit{t}) (right \textit{y}-axis) of transmission coefficient while only Q\textit{i} is resonant. The blue (orange) dots represent the measured magnitude (phase). The black solid curves denote the fitting results. }
\label{fig:singleQB_all}
\end{figure}

We characterize single-qubit performance by scattering experiments of each qubit. 
As shown in Fig.~\ref{fig:singleQB_all}, we measure the transmission spectrum of each qubit while keeping other qubits far-detuned. 
We fit the complex transmission spectra with following equation of $t$ \cite{Hoi_2013}: 
\begin{equation}
	t=1-\frac{\Gamma_i}{\Gamma_i+\gamma_i'}\cdot\frac{1-{\rm i}\delta_{\rm p}\frac{\Gamma_i+\gamma_i'}{2}}{1+\left(\frac{2\delta_{\rm p}}{\Gamma_i+\gamma_i'}\right)^2}.
	\label{eq:t}
\end{equation}
Here, $\delta_\text{p}=\omega_{\text{p}}-\omega_{\text{0}}$ is the detuning between the probe and qubit frequency. Circular fitting method is used to acquire radiative decay rates into waveguide $\Gamma_i$ and non-radiative dissipation rates $\gamma_i'$. The measured \textit{t} spectrum and the fitting results in Devices A, B1 and B2 are shown in Fig. \ref{fig:singleQB_all}. Table \ref{tab.singleQB_all} shows the extracted $\Gamma_i$ and $\gamma_i'$ of each qubit. 

\begin{table}[h]
	\caption{Extracted radiative decay rates into waveguide $\Gamma_i$ and non-radiative dissipation rates $\gamma_i'$ of the qubits in Devices A, B1 and B2 (unit: MHz).}
	\label{tab.singleQB_all}
	\centering
	\begin{tabularx}{0.6\textwidth} {
			 >{\raggedright\arraybackslash}X 
			 >{\raggedleft\arraybackslash}X 
			 >{\centering\arraybackslash}X 
			 >{\centering\arraybackslash}X 
			 >{\centering\arraybackslash}X 
			 >{\centering\arraybackslash}X 
			 >{\centering\arraybackslash}X 
			 >{\centering\arraybackslash}X 
			 >{\centering\arraybackslash}X 
			 >{\centering\arraybackslash}X 
			}		
		\hline
		&                  & Q1     & Q2    & Q3    & Q4     & Q5     & Q6    & Q7     & Q8     \\ \hline
		\multirow{2}{*}{Device A}  & $\Gamma_i/2\pi$  & 10.132 & 9.759 & 9.738 & 10.172 & 10.049 & 9.824 & 10.931 & 10.084 \\
		& $\gamma_i'/2\pi$ & 0.304  & 0.313 & 0.268 & 0.290  & 0.328  & 0.283 & 1.096  & 0.283  \\
		\multirow{2}{*}{Device B}  & $\Gamma_i/2\pi$  & 7.834  & 5.664 & 7.668 & 6.495  & 7.255  & 6.868 & 6.807  & 6.791  \\
		& $\gamma_i'/2\pi$ & 0.643  & 1.108 & 0.174 & 1.889  & 0.429  & 1.048 & 0.513  & 1.179  \\
%		\multirow{2}{*}{Device B2} & $\Gamma_i/2\pi$  &        & 5.611 & 6.458 & 5.848  & 6.273  & 6.059 & 6.215  & 5.688  \\
%		& $\gamma_i'/2\pi$ &        & 0.649 & 0.301 & 0.761  & 0.336  & 0.733 & 0.586  & 0.724  \\ 
		\hline
	\end{tabularx}
\end{table}

\clearpage

\clearpage

\section{transfer matrix method}\label{sec:transfer_matrix}

To calculate the transmission and reflection spectrum, we utilize the transfer matrix method\cite{PhysRevA.100.013812}. The total transfer matrix through the waveguide can be expressed by multiplying the matrices of the qubits and the waveguide of length $d$: 
\begin{equation}
	\label{eq.transfer_matrix}
	M_{\text{total}}=\prod_{m=1}^{N}M_{m,\text{qubit}}M_{\text{waveguide}}. 
\end{equation}
Here $M_{m,\text{qubit}}$ denotes the transfer matrix of the \textit{m}th qubit depending on the transmission and reflection coefficient ($t$ and $r$). $M_{\text{waveguide}}$ represents the matrix of the waveguide between qubits, indicating that the photon acquires phase when propagating through the waveguide. The expressions of both matrices are shown below: 
\begin{equation}
	M_{m,\text{qubit}} = \frac{1}{t_m}
	\left( {\begin{array}{cc}
			t_m^2-r_m^2 & r_m   \\
			-r_m        & 1     \\
	\end{array} } \right), 
\end{equation}

\begin{equation}
	M_{\text{waveguide}} = 
	\left( {\begin{array}{cc}
			e^{ik_xd} & 0 \\
			0 & e^{-ik_xd} \\
	\end{array} } \right). 
\end{equation}
The reflection coefficient $r_m$ of the $m$th single qubit can be obtained from the definition $t_m=1+r_m$. Here $t_m$ is the transmission coefficient taking the form of Eq. \ref{eq:t}. The reflection coefficient $r_m$ writes:
\begin{equation}\
	r_m=-\frac{\Gamma_i}{\Gamma_i+\gamma_i'}\cdot\frac{1-{\rm i}\delta_{\rm p}\frac{\Gamma_i+\gamma_i'}{2}}{1+\left(\frac{2\delta_{\rm p}}{\Gamma_i+\gamma_i'}\right)^2}.
\end{equation}

By calculating Eq. \ref{eq.transfer_matrix}, the total transmission $t$ and reflection coefficient $r$ can be obtained: 
\begin{equation}
	r=-\frac{[M_{\text{total}}]_{2,1}}{[M_{\text{total}}]_{2,2}},~
	t=\frac{\det M_{\text{total}}}    {[M_{\text{total}}]_{2,2}}.~
\end{equation}

\clearpage
\section{bandwidth of the anti-Bragg bandgap}

\begin{figure}[h]
	\centering
	\includegraphics[]{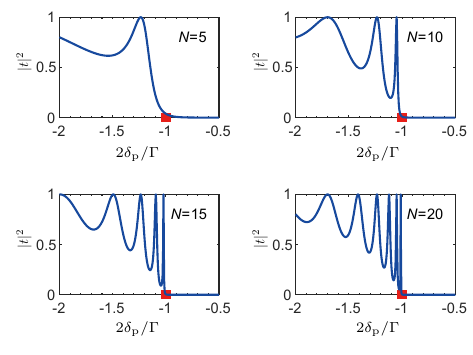}
	\caption{The calculated $|t|^2$ spectra of anti-Bragg spaced qubit arrays. Different conditions of $N$=5, 10, 15 and 20 are considered. The red square denotes the approximate boundary of the flat bottom at $\delta_\text{p}=-\Gamma/2$. }
	\label{smfig.bandwidth}
\end{figure}

Based on the transfer matrix method, the transmittance $|t|^2$ and the reflectance $|r|^2$ of a $N$ resonant qubit array near a waveguide can be written in the following form\cite{PhysRevA.103.023508, PhysRevA.100.013812}: 
\begin{equation}
	|t|^2=
	\frac{1}
	{1+\alpha^2U_{N-1}^2(y)},
	|r|^2=
	\frac{\alpha^2U_{N-1}^2(y)}
	{1+\alpha^2U_{N-1}^2(y)}
\end{equation}
where $\alpha=\Gamma/2\delta_{\rm p}$ and
% $y=\cos(kd)+\alpha\sin(kd)$.
$y=\cos(\theta)+\alpha\sin(\theta)$. Here we consider the atoms are homogeneous in radiative decay rate into waveguide, i.e, $\Gamma_i=\Gamma$. 
And $U_{N-1}^2(y)$ is the Chebyshev polynomials of the second kind. Near the resonance $\omega_\text{0}$, the transmission is strongly suppressed within the band $|y|>1$. We write $\theta$ as 
\begin{equation}
	\theta=\frac{\omega_{\text{p}}}{c}d=k_0d(1+\frac{\delta_{\text{p}}}{\omega_\text{0}}). 
\end{equation}
At the anti-Bragg condition, we have $k_0d=\pi/2$. Assuming $\delta\omega_{\text{p}}\ll\omega_{\text{0}}$, the approximate boundaries of the band $|y|>1$ are $\delta_{\text{p}}\approx\pm\Gamma/2$. This explains the width of the flat bottom of $|t|^2$ spectrum is $\Gamma$ and independent of $N$. We also show the calculated $|t|^2$ spectra under different conditions of $N$=5, 10, 15 and 20 in Fig. \ref{smfig.bandwidth}. The calculated results support the weak dependence of the flat bottom's width on qubit number $N$.

\clearpage
\section{calculation for the scenario of introducing single-qubit detuning}

\begin{figure}[h]
	\centering
	\includegraphics[]{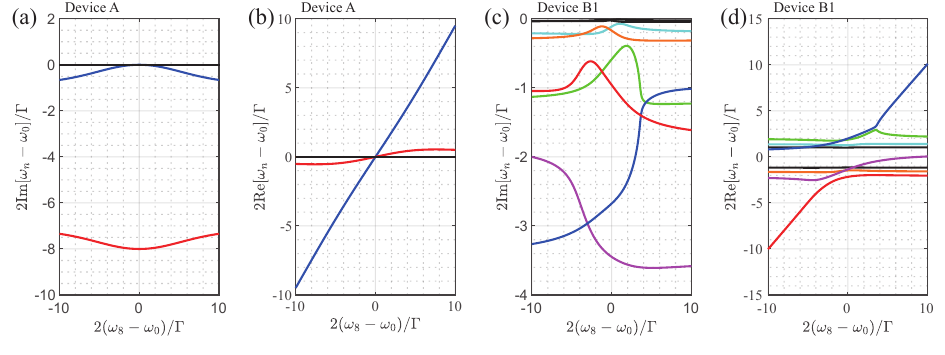}
	\caption{Calculated eigenstates when Q8's frequency $\omega_8$ is tuned across the resonance. 
		(a), (b) Detuning and radiative decay rate of the eigenstates for Device A. 
		(c), (d) Detuning and radiative decay rate of the eigenstates for Device B1. 
		The calculation is performed in the absence of non-radiative decay. }
	\label{smfig.q8_eigen}
\end{figure}

We calculate the eigenstates of eight-qubit system while detuning the last qubit Q8 through the resonance. Fig \ref{smfig.q8_eigen} shows the calculated eigenstates in Devices A and B1. The calculation is performed in the absence of non-radiative dissipation rate. 
In Fig \ref{smfig.q8_eigen}(a) and (b), the black curve represent the six dark states, the red and blue curve denote the eigenstates resulted from waveguide-mediated interaction between Q8 and the bright state of the $N=7$ Bragg superatom. Fig \ref{smfig.q8_eigen}(c) and (d) show the eigenstates in the anti-Bragg condition. The eigenstates are radiative and respond to Q8's detuning differently. Interference between these states could result in complicated scattering spectrum. 

\begin{figure}[h]
	\centering
	\includegraphics[]{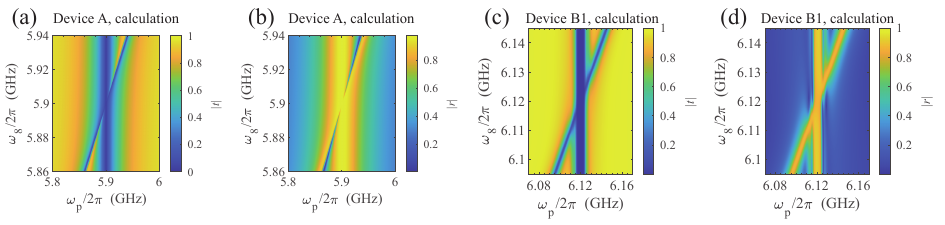}
	\caption{Calculated scattering spectrum when Q8's frequency $\omega_8$ is tuned across the resonance. 
		(a), (b) Calculated $|t|$ and $|r|$ for the measurement in Device A. The resonance frequency of Q1-Q7 is $\omega_0/2\pi=$5.9 GHz. 
		(c), (d) Calculated $|t|$ and $|r|$ for the measurement in Device B1. The resonance frequency of Q1-Q7 is $\omega_0/2\pi=$6.12 GHz. 
%		The measured $\Gamma_\text{r}$ and $\Gamma_{2}$ of individual qubits are used in the calculation. 
		}
	\label{smfig.q8_simspectrum}
\end{figure}

Then we calculated the scattering spectrum for the measurement of detuning Q8 in Devices A and B1, as Fig \ref{smfig.q8_simspectrum} shows. Here, we use transfer matrix method (for detail see Sec. \ref{sec:transfer_matrix}) and extracted $\Gamma_i$ and $\gamma_i'$ of individual qubits in Table \ref{tab.singleQB_all} for the calculation. 

\clearpage
\section{observed multiple subradiant states near Bragg condition}

\begin{figure}[h]
	\centering
	\includegraphics{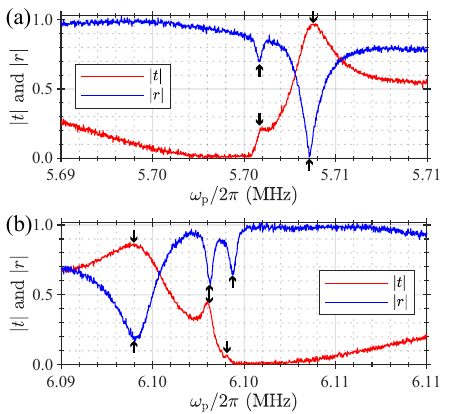}
	\caption{Measured transmission (red) and reflection (blue) amplitudes when (a) $\omega_0/2\pi=$ 5.700 GHz and (b) $\omega_0/2\pi=$ 6.100 GHz of the Bragg superatom in Device A. The arrows indicate the multiple transparency windows in the measured spectra. }
	\label{smfig.vsRatio_Fine}
\end{figure}

For a Bragg superatom, as the resonance frequencies of the array deviate from the Bragg condition, the dark states become subradiant states which are not degenerate with the superradiant state. Quantum interference between these states can induce Fano resonances. As the Fig. 1(d) in the main text shows, there are 7 subradiant states for a size \textit{N}=8 Bragg superatom. Figure \ref{smfig.vsRatio_Fine} show the measured transmission and reflection amplitudes when $\omega_0/2\pi=$ 5.700 GHz and $\omega_0/2\pi=$ 6.100 GHz of the Bragg superatom in Device A, corresponding to $d/\lambda_0\approx0.483$ and $d/\lambda_0\approx0.517$. Here, we observe multiple transparency windows induced by the interference effect. However, only up to 3 brightest subradiant states are observed in Fig. \ref{smfig.vsRatio_Fine}(b) while the other subradiant states with narrower linewidth are not observed due to decoherence. 

% \bibliographystyle{}
\bibliography{sm_reference}